\documentclass[aps, prb ,twocolumn,superscriptaddress]{revtex4-1}

\usepackage{float}
\usepackage{xcolor}
\usepackage{graphicx}
\usepackage[dvips]{epsfig}
\usepackage{bm, amssymb}
\usepackage{dcolumn}

\begin{document}

\title{Magnetic correlations in subsystems of the misfit [Ca$_2$CoO$_3$]$_{0.62}$[CoO$_2$] cobaltate}
\author{Abdul Ahad}
\address{Department of Physics, Aligarh Muslim University, Aligarh 202002, India}
\author{K. Gautam}
\address{UGC-DAE Consortium for Scientific Research, Indore 452001, India}
\author{K. Dey}
\address{UGC-DAE Consortium for Scientific Research, Indore 452001, India}
\author{S. S. Majid}
\thanks{Present address: Optical Physics Lab, Institute of Physics, Academia Sinica, Taipei, Taiwan.}
\address{Department of Physics, Aligarh Muslim University, Aligarh 202002, India}
\author{F. Rahman}
\address{Department of Physics, Aligarh Muslim University, Aligarh 202002, India}
\author{S. K. Sharma}
\address{Department of Physics, Central University of Punjab, Bathinda Punjab 151001, India}
\author{J. A. H. Coaquira}
\address{LSNCM-NFA, Institute of Physics, UnB, Brasilia DF 70910 900, Brazil}
\author{Ivan da Silva}
\address{ISIS Facility, Rutherford Appleton Laboratory, Chilton, Didcot OX11 0QX, United Kingdom}
\author{E. Welter}
\address{Deutsches Elektronen-Synchrotron, Notkestrasse 85, D-22607 Hamburg, Germany}
\author{D. K. Shukla}
\thanks{Corresponding Author: dkshukla@csr.res.in}
\address{UGC-DAE Consortium for Scientific Research, Indore 452001, India}


\date{\today}

\begin{abstract}
[Ca$_2$CoO$_3$]$_{0.62}$[CoO$_2$], a two dimensional misfit metallic compound, is famous for its rich phases accessed by temperature, $i.e.$ high temperature spin-state transition, metal-insulator transition (MIT) at intermediate temperature ($\sim$ 100 K) and low temperature spin density wave (SDW). It enters into SDW phase below T$_{MIT}$ which becomes long range at 27 K. Information on the independent role of misfit layers (rocksalt/Ca$_2$CoO$_3$ \& triangular/CoO$_2$) in these phases is scarce. By combining a set of complementary macroscopic (DC magnetization and resistivity) and microscopic (neutron diffraction and X-ray absorption fine structure spectroscopy) measurements on pure (CCO) and Tb substituted in the rocksalt layer of CCO (CCO1), magnetic correlations in both subsystems of this misfit compound are unraveled. CCO is found to exhibit glassiness, as well as exchange bias (EB) effects, while CCO1 does not exhibit glassiness, albeit it shows weaker EB effect. By combining local structure investigations from extended X-ray absorption fine structure (EXAFS) spectroscopy and neutron diffraction results on CCO, we confirm that the SDW arises in the CoO$_2$ layer. Our results show that the magnetocrystalline anisotropy associated with the rocksalt layer acts as a source of pinning, which is responsible for EB effect. Ferromagnetic clusters in the Ca$_2$CoO$_3$ affects SDW in CoO$_2$ and ultimately glassiness arises.
\end{abstract}

\pacs{kk}  \maketitle

Magnetism in misfit cobaltates is a debated topic of investigation although interesting~\cite{Bay2005}. The misfit structure makes the physics of these systems complex. For example, a famous misfit structure, sodium cobaltate (Na$_x$CoO$_2$), offers superconductivity in hydrated form and thermoelectricity with the metallic conductivity~\cite{Tak2003,Ter1997}. Moreover, the existence of cobalt ion (having spin state variants~\cite{Abd2017}) in such misfit cobaltates makes the task daunting for the magnetic structure prediction. Besides, another ingredient of complexity is geometric frustration due to triangular lattice CoO$_2$, having edge shared Co ions octahedra in D$_{3d}$ symmetry~\cite{Sen2007}. In Na$_x$CoO$_2$, sodium content decides the valency of Co ions in the triangular lattice (CoO$_2$) and it shows rich phases with different concentrations of Na, $e.g.$, the extreme member, Na$_x$CoO$_2$ (x = 1) is a non-magnetic insulator~\cite{Rav2012} and for x $\sim$ 0.62, the compound shows the boundary in between the anti-ferromagnetic (AFM) and ferromagnetic (FM) correlations dominant compositions~\cite{Lan2008}. In the crystal structure of Na$_x$CoO$_2$, the CoO$_2$ layers are separated by the layers of Na atoms and even with the two dimensional structure it has been found that for such structures interlayer and intralayer magnetic interaction have comparable strength~\cite{Bay2005}. 

 Famous for its thermoelectricity the Ca$_3$Co$_4$O$_9$, more precisely [Ca$_2$CoO$_3$]$_{0.62}$[CoO$_2$] (hereinafter abbreviated as CCO), has two subsystems as intergrowth of one on the other aperiodically. According to the chemical formula, it is comparable with x $\sim$ 0.6 composition of Na$_x$CoO$_2$. One can roughly compare the magnetism of the CoO$_2$ layer in both structures, however, in CCO the role of the [Ca$_2$CoO$_3$] layer (having stack of CaO-CoO-CaO with rocksalt structure) is significant, therefore the overall magnetic behavior is unique.
 
  CCO exhibits the onset of SDW below $T_{MIT}$ $\sim$ 100 K which become long range at $T_{SDW}$ $\sim$ 27 K followed by ferrimagnetic ordering at $T_{Ferri}$ $\sim$ 19 K. Many researchers have tried to alter its properties by doping. For example, it is reported that Sr doping at the Ca site weakens the ferrimagnetism and shows AFM correlations~\cite{Sug2002}. The electron doping at the Co site of the rocksalt layer by the trivalent ion doping at Ca site (Y$^{3+}$ \& Bi$^{3+}$) diminishes the ferrimagnetism and affects the  $T_{SDW}$ which highlights the role of Co valency in the rocksalt layer~\cite{Sug2002}. It is also reported that the SDW in the CoO$_2$ subsystem has oscillating moments in the $c$ direction and motion in the $ab$ plane and, by comparing the results with the doped CCO, it is suggested that SDW can be tuned by doping in the rocksalt layer at the Ca site~\cite{Sugi2003}.
 
 \begin{figure}[hbt]
 	\centering
 	\includegraphics[width=0.46\textwidth]{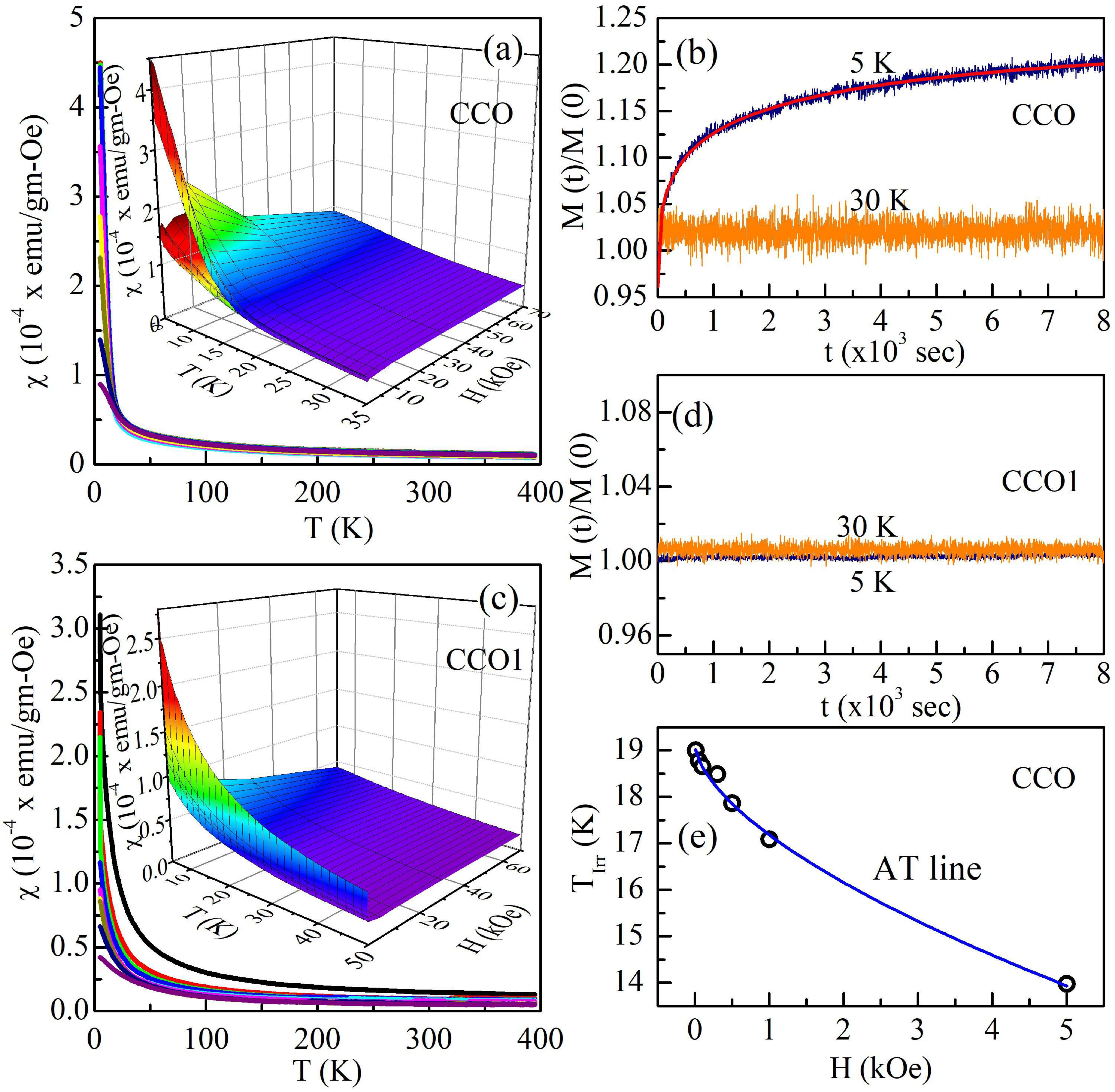}
 	\caption{Field cooled susceptibility as a function of temperature, measured at different magnetic fields (a) for CCO and (c) for CCO1. Insets show the 3D plot ($\chi$-H-T), showing the bifurcation between FC and ZFC. ZFC relaxation curves measured at 5 K and 30 K under the same magnetic field (50 Oe), (b) for CCO and (d) for CCO1. (e) Dependence of T$_{Irr}$ on magnetic field for CCO. Blue solid line shows the fitting to data (see text).}
 	\label{1}
 \end{figure}

\begin{figure}[hbt]
	\centering
	\includegraphics[width=0.48\textwidth]{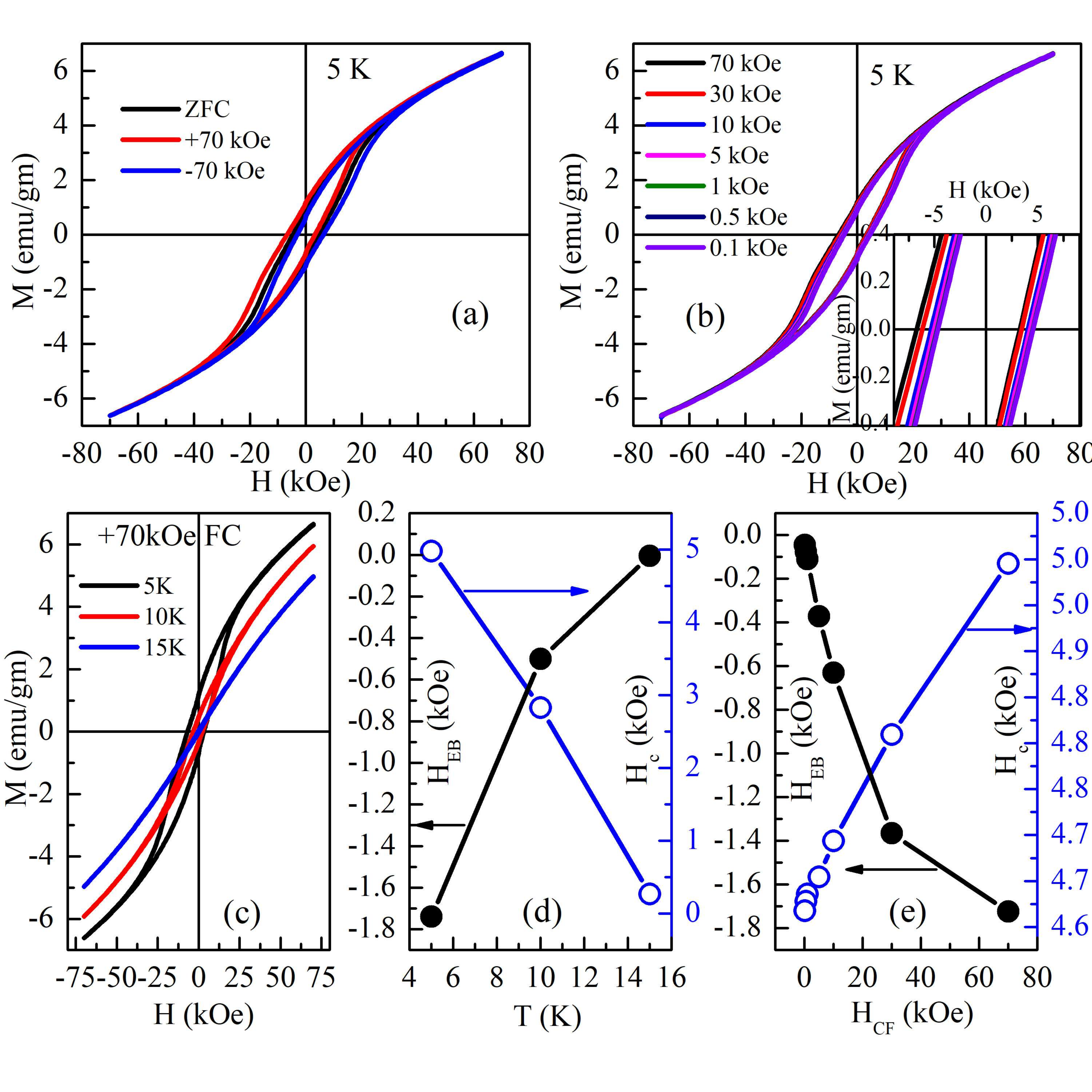}
	\caption{For CCO, (a) M(H) isotherm loops measured at 5 K in ZFC and FC (+70 kOe and -70 kOe), showing the hysteresis and EB (in field cooled cases). (b) M(H) loop at 5 K measured after cooling under various fields. Inset shows the successive shifting of loop on the field axis. (c) M(H) loops measured at different temperatures in +70 kOe field cooled condition. (d) Temperature dependence of H$_{EB}$ and H$_c$. (e) Cooling field dependence of H$_{EB}$ and H$_c$.}
	\label{2}
\end{figure}

Here we report on the drastic alteration in magnetic properties of the CCO by electron doping at the Co site in the rocksalt layer by Tb doping at the Ca site. Doping concentration of Tb is decided on the basis of earlier studies of CCO~\cite{Sai2017}. We have utilized the exchange bias, present in both CCO and Tb substituted CCO, as a tool to discern the role of different magnetic lattices. Competition between the rocksalt layer $c$ axis magnetism and triangular layer itinerant magnetism has been found as the cause of ferrimagnetism. Magnetocrystalline anisotropy associated with the rocksalt layer has been identified as the cause of pinning for exchange bias. Concomitant with the broadness of neutron diffraction peaks, an anomaly in the spin phonon coupling in the CoO$_2$ layer (observed via EXAFS) confirms the truncation of long range incommensurate SDW (ISDW) into a glassy phase. The Tb doping has been found to change the effect of rocksalt on the SDW in CoO$_2$ by screening the rocksalt field, and consequently CCO1 shows no glassiness and less EB in CCO1.

Pure (CCO) and Tb doped Ca$_{2.9}$Tb$_{0.1}$Co$_4$O$_9$ (CCO1) have been synthesized using solid state route, as reported elsewhere~\cite{Kan2014}. Phase purity of the samples have been confirmed using X-ray diffraction~\cite{Abd2020}. X-ray photoemission spectroscopy (XPS) has been performed using an Omicron energy analyzer (EA-125) with Al $K\alpha$ (1486.6 eV) X-ray source. Magnetization measurements were done using a 7T Quantum Design magnetometer (MPMS-3). Isotherms, virgin and full loop M (H) have been recorded at various temperatures across the mentioned transitions, $i.e.$, T$_{SDW}$ and T$_{Ferri}$ in FC and ZFC modes. Magnetization as a function of temperature M (T) at different applied magnetic fields were recorded in FC and ZFC protocols. ZFC relaxation measurements have been done at 5 and 30 K by cooling the sample in zero field down to the desired temperature and, after a 100 s delay, magnetization have been recorded at 50 Oe for upto 8000 s. Neutron diffraction patterns have been collected at General Materials Diffractometer (GEM), ISIS facility, UK, in the temperature range 6-110 K. JANA 2006~\cite{Dus2006} was used for fitting the neutron diffraction patterns. Extended X-ray absorption fine structure spectroscopy (EXAFS) measurements have been performed at beamline P65 at PETRA III, DESY, Germany. The EXAFS measurements were done in fluorescence and transmission mode at Co K edge (7.7 keV). The sample amount was calculated for one absorption length and homogeneously mixed with boron nitride and pressed into a pellet shape. A liquid helium flow cryostat has been used for low temperature EXAFS measurements. $Athena$ has been utilized for data processing. In $Artemis$, the FEFF and IFEFFIT codes were used to calculate theoretical scattering paths and to fit the experimental spectra, respectively.

First we will discuss the results of CCO. FC magnetic susceptibility ($\chi$) (see Fig.~\ref{1} (a)) with the bifurcation in FC and ZFC in low field (see inset) indicates presence of magnetic glassiness of some type or the presence of magnetocrystalline anisotropy or both together~\cite{Joy1998}. The upturn in the $\chi$ has been attributed as T$_{Ferri}$, in literature, while in magnetic entropy ($\Delta S_M$) it is visible as a first derivative of M at $\sim$ 10 K (see Fig. S1 (a) in the supplementary file~\cite{Sup}). Also, there is no spontaneous magnetization observed via Arrott plot at all measurement temperatures (see Fig. S2 (a) of supplementary~\cite{Sup}). ZFC relaxation measurements (see Fig.~\ref{1} (b)) show the time dependence of magnetization at 5 K but not at 30 K. To confirm the glassiness~\cite{Abd2019} we have fitted the ZFC relaxation curve with the stretched exponential function $M (t) = M_o - M_r exp(-(t/t_r)^\beta)$ where the value of $\beta$ tells the distribution of barrier and is found to be $\sim$ 0.37, which is close to the value for canonical spin glass~\cite{Abd2019} ($\sim$ 0.42). From the inset of Fig.~\ref{1} (a) it is observed that the bifurcation exists below $\sim$ 20 K for magnetic fields up to $\sim$ 40 kOe. The bifurcation temperature (T$_{Irr}$) followed a trend with magnetic field, unique for spin-glass (see Fig.~\ref{1} (e)). Fitting with the equation $T_{Irr}(H)=T_{Irr}(0)(1-AH^n)$ reveals the exponent n $\sim$ 0.66, which is typical for Almeida-Thouless (AT) line, predicted theoretically for spin-glass\cite{Bin1986,Pak2016}. Based on preliminary analysis for now, we designate the bifurcation as related to the glassiness.

\begin{figure}[t]
	\centering
	\includegraphics[width=0.48\textwidth]{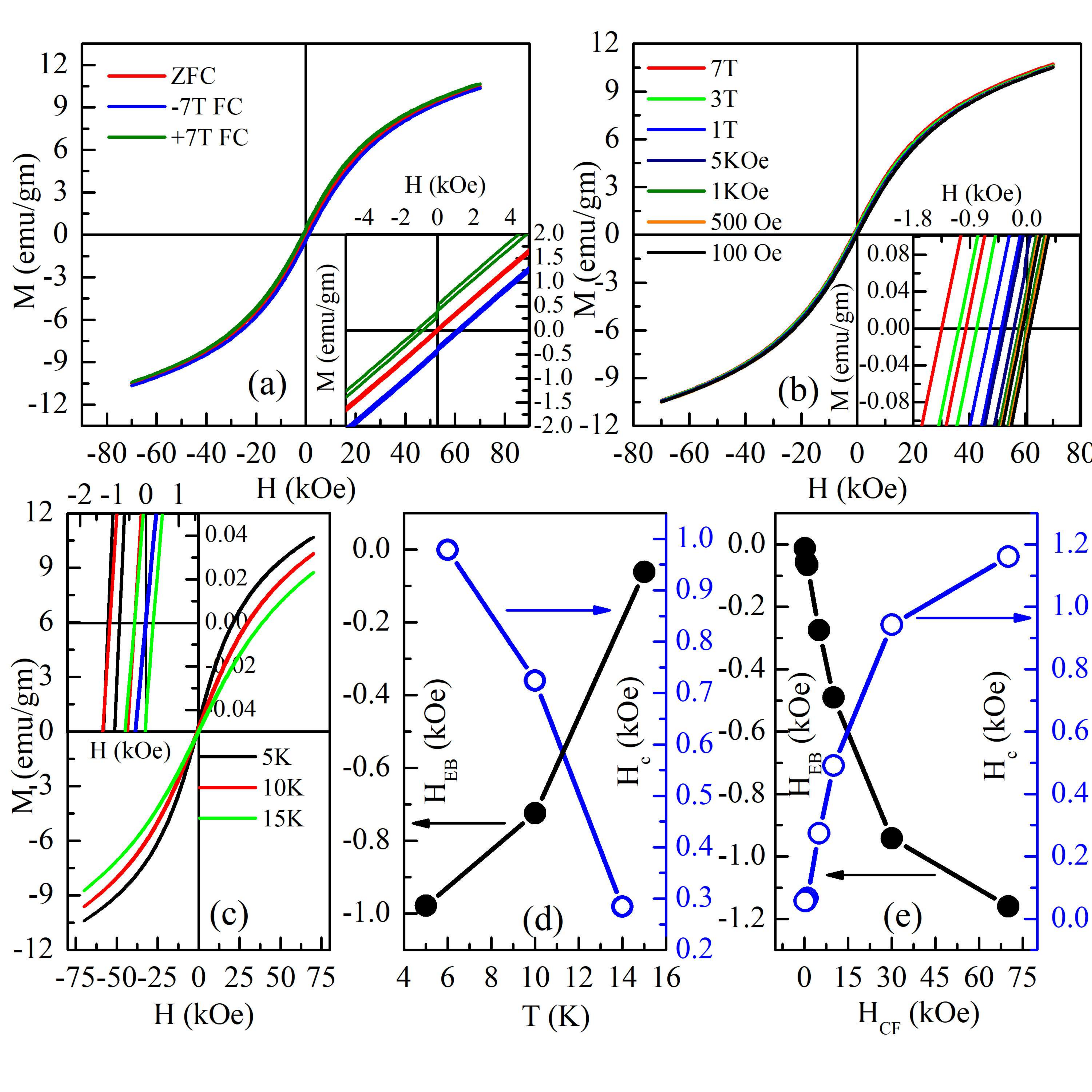}
	\caption{For CCO1, (a) M(H) isotherm loop measured at 5 K in ZFC and FC (+70 kOe and -70 kOe), showing the hysteresis (see inset) and EB (in field cooled cases). (b) M(H) loop at 5 K measured after cooling under various fields. Inset shows the successive shifting of loop on the field axis. (c) M(H) loops measured at different temperatures in +70 kOe field cooled condition. (d) Temperature dependence of H$_{EB}$ and H$_c$. (e) Cooling field dependence of H$_{EB}$ and H$_c$.}
	\label{4}
\end{figure}

Interestingly, we have observed the exchange bias (EB) in CCO at 5 K, +70 kOe FC with magnitude H$_{EB}$ $\sim$-1.7 kOe and coercivity  H$_{c}$ $\sim$ 5 kOe, calculated using $H_{EB} = \left(H_{c1} + H_{c2}\right)/2$ and $H_{c} = \left(|H_{c1}| + |H_{c2}|\right)/2$, respectively. Here H$_{c1}$ and H$_{c2}$ are the coercive fields in the negative and positive field side, respectively. The magnitude is considerably large, however one has to authenticate the existence of it. Fig.~\ref{2} (a) shows the M (H) hysteresis measured in ZFC, +70 kOe FC and -70 kOe FC at 5 K. The loop shifted to negative and positive directions for cooling in positive and negative fields, respectively. This is according to conventional EB system~\cite{Nog1999}. The cooling field dependence (see Fig.~\ref{2} (b)) at 5 K and temperature dependence at +70 kOe has been observed (see Fig.~\ref{2} (c)). These trends also match with the conventional EB cases~\cite{Nog1999,Tan2006} (see Fig.~\ref{2} (d \& e)).

For conventional EB systems with AFM and FM layers with the strong interfacial coupling, the H$_{EB}$ is defined as~\cite{Mei1956,Kar2008} $H_{EB} = -J \frac{S_{AFM} S_{FM}}{\mu_o t_{FM} M_{FM}}$ where $J$ represents the coupling strength across the interface, $S_{FM/AFM}$ is the interface magnetization of FM/AFM phase, and $t_{FM}$ \& $M_{FM}$ are the thickness and bulk magnetization of the FM layer. From this relation, it is clear that H$_{EB}$ will increase with the increase in interfacial FM, which increases with the cooling field (H$_{CF}$) due to spin alignment in field direction. Although, enhancement in the H$_{CF}$ results in increase in cluster size (decreases $S_{FM}$) and enhances the bulk magnetization, $M_{FM}$ therefore reducing~\cite{Kar2008} the H$_{EB}$. Moreover, for the phase separated systems~\cite{Tan2006}, with FM clusters in the SG matrix, the above situation is also observed, but the effect of magnetic field on the glassy phase has to be considered, which usually diminishes with the applied field.

\begin{figure}[hbt]
	\centering
	\includegraphics[width=0.45\textwidth]{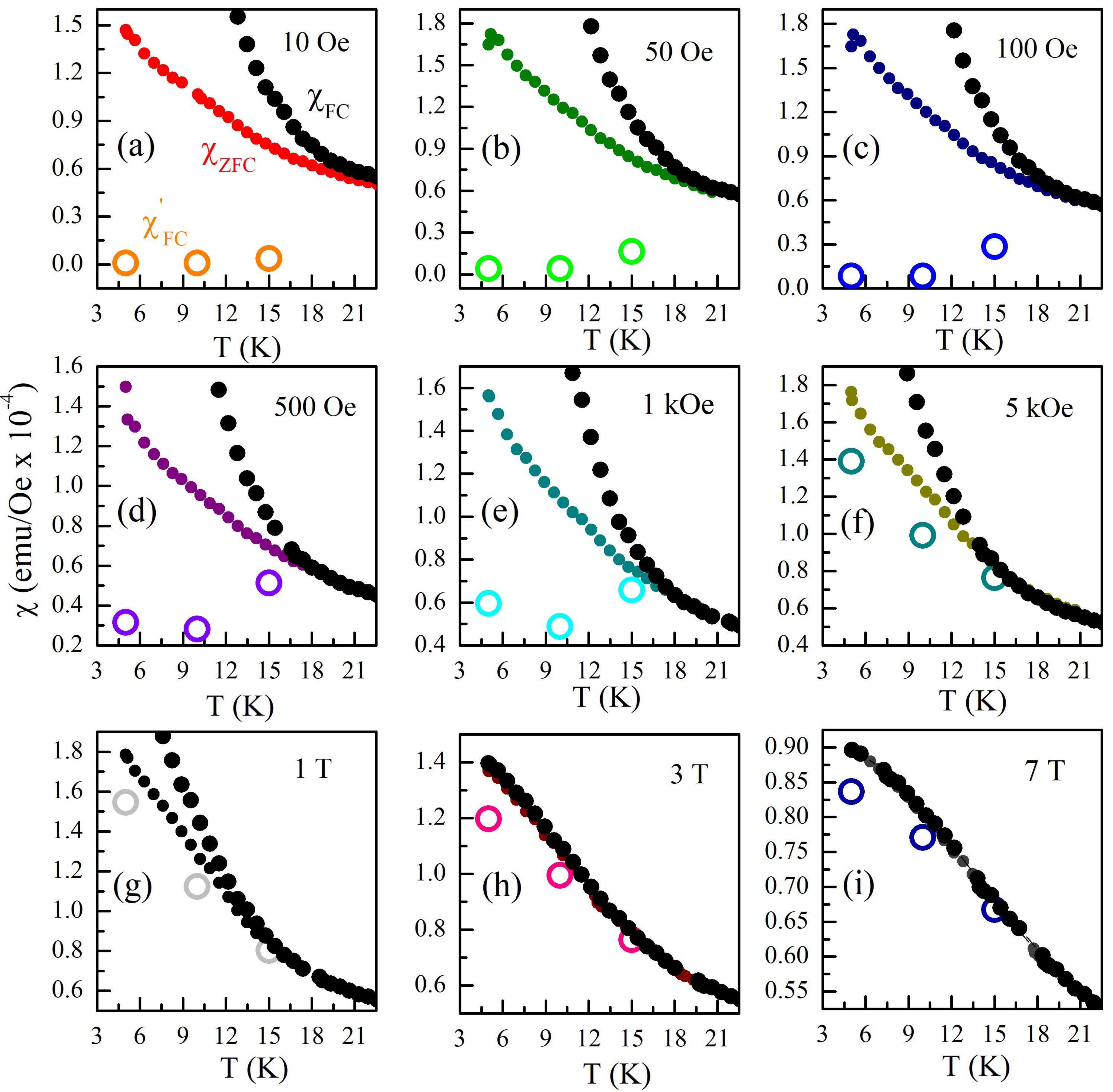}
	\caption{(a)-(i) Comparison between $\chi_{FC}$ (black symbols), $\chi_{ZFC}$ (filled symbols) and calculated $\chi'_{FC}$ (empty circles) for the CCO (see text).}
	\label{7}
\end{figure}

The important and unusual observations in the present case are the suppression of EB (Fig.~\ref{2} (d)) and bifurcation (Fig.~\ref{1} (a)) for temperature $\gtrsim$ 15 K, the non saturation behavior of H$_{EB}$ up to 70 kOe (Fig.~\ref{2} (e)) and the suppression of bifurcation in a field above $\sim$ 40 kOe (see Fig.~\ref{7} and related texts). These open the question about the origin of EB, because if the glassiness is considered as the origin of pinning, then it should vanish for a field above $\sim$ 40 kOe, which is not the case here.

 Before making any comment on the origin of EB, we will discuss the results of CCO1 (Tb doped CCO). For the doping at the Ca site, the chemical formula can be written as [Ca$_{1.959}$Tb$_{0.041}$CoO$_3$]$_{0.62}$[CoO$_2$]. Tb$^{3+}$ is a magnetic ion with the total spin moment S = 3 ($4f^8$) with the theoretical Ising moment\cite{Szy2007,Rev} $m_z$ $\sim$ 9.72 $\mu_B$. Figs.~\ref{1} (c) and (d) display the $\chi$(T) and isothermal ZFC relaxation, respectively. The magnitude of moment is larger for CCO1 because of the additional paramagnetic contribution from the Tb. No transition of any type is observed in the $\Delta S_M$ (see Fig. S1 (b)~\cite{Sup}), which means that the Tb destabilizes the ferrimagnetism. Arrott plot is similar to that for CCO, $i.e.$ M$_s$ = 0 (see Fig. S2 (b) of supplymentry~\cite{Sup}). These observations indicate that there is no glassiness in the CCO1 and it should not behave like CCO, $i.e.$, it should not possess EB. We have carried out the same set of magnetization measurements as done for CCO, shown in Fig.~\ref{4} (a-e). Counterintuitive, this system, CCO1, also exhibits the EB albeit with lower strength (H$_{EB}$) and with lower coercivity (H$_c$). In H$_{EB}$ plot as a function of cooling field H$_{CF}$, no saturation or decreasing trend has been observed up to 70 kOe, similar to CCO.

In order to know what is happening at microscopic scale, photoemission spectroscopy measurements were carried out on both the samples which clearly indicates that the Tb substitution increases the Co$^{3+}$ (see Fig. S4~\cite{Sup}). A similar observation has been made by other groups for high concentration of Tb doping~\cite{Sai2017}.

 The observations from CCO1 indicate that the origin of bifurcation in CCO may come from the magnetocrystalline anisotropy and therefore it should follow the equation suggested by Joy $et~al.$~\cite{Joy1998,Abd2019} given as $\frac{M_{FC}}{H_{appl} + H_c}$ (= $\chi'_{FC}$) $\approx$ $\frac{M_{ZFC}}{H_{appl}}$ (= $\chi_{ZFC}$). Figs.~\ref{7} (a-i) show that large bifurcation appears because of the glassy phase, which is suppressed for higher fields ($>$ 5 kOe), and for higher field the magnetocrystalline anisotropy model is satisfied. This explains the reason for large hysteresis in CCO, $i.e.$, glassy phase, while CCO1 does not exhibit it. However, the existence of EB in both samples indicates that the origin of EB is not the glassiness. Therefore, it is proposed that the Tb doping affects the magnetocrystalline anisotropy, which indirectly suppresses the glassiness. For, ingredient of EB~\cite{Nog1999}, we have AFM SDW in the CoO$_2$ layer~\cite{Sug2003}, which is common in both the samples. 
 
\begin{figure}[hbt]
	\centering
	\includegraphics[width=0.48\textwidth]{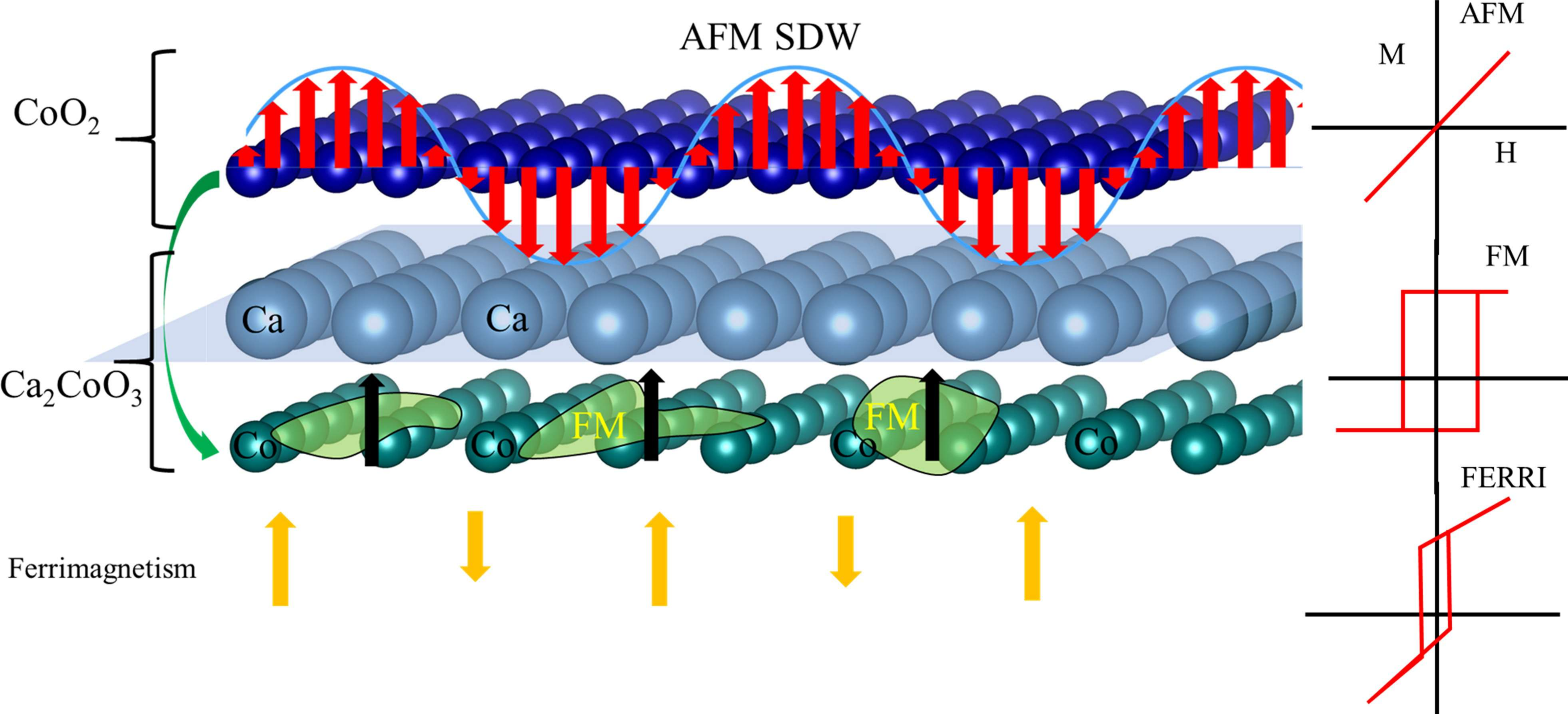}
	\caption{Schematic showing the AFM SDW in the CoO$_2$ layer, having sinusoidal behavior (red arrows), resulting in a linear M(H) behavior. In the rocksalt layer the green clusters show short range ferromagnetic arrangement of spins with strong spin orbit coupling (SOC), which restricts the magnetism in the crystal $c$ axis, resulting in a hysteretic M(H) loop. The resultant of these two gives rise to overall ferrimagnetism (orange arrows), resulting in a non-saturating hysteretic M(H). The layer of Ca in between these two magnetic layers controls the coupling.}
	\label{8}
\end{figure}

 The SDW  generally appears as an AFM ordering in low dimensional metallic systems. This is simple fact by which one can distinguish the AFM (localized insulating) and SDW (metallic systems). SDW is itinerant but can show the similar behavior as showed by localized helical or cycloidal ordered systems. These orderings can be described by the orientation of spins S in all directions (S$_x$, S$_y$ and S$_z$) having same magnitude $\mid S \mid$, while for the SDW the direction remains the same (S$_x$, S$_y$ or S$_z$) but the magnitude has an oscillatory behavior~\cite{Eli2018} ($Scos(2q.r)$). In CCO the rocksalt layer possesses short range FM because of clustering of mixed valency (Co$^{3+}$ and Co$^{4+}$)~\cite{Abd2020}. And the low spin state (LSS) Co$^{4+}$ (S = 1/2, $\tilde{L}$=1) can offer the magnetocrystalline anisotropy through spin-orbit coupling (SOC)~\cite{Csi2005,Hol2008}. In CCO1, as a result of Tb doping at the Ca site, the interlayer coupling ($J_\perp$) decreases, as well as the amount of Co$^{4+}$ in the rocksalt layer. This can explain the low temperature shifting of T$_{Ferri}$ or its absence, as a result of the decrease in the $J_\perp$ value associated with the interlayer coupling ($J_\perp~\propto~k_B T$). The schematic shown in Fig.~\ref{8} represents the above mentioned hypothesis that CoO$_2$ exhibits AFM SDW with spins arranged in a wave pattern (red arrow). In the rocksalt layer the FM clusters (made up of Co$^{4+}$ - Co$^{3+}$) provide the short range ferromagnetism along the $c$ axis; black arrows shows their effective strength and direction. The effective moments, as a sum of these two contributions, result into overall ferrimagnetism. The Ca layer in between the CoO$_2$ and Ca$_2$CoO$_3$ controls their coupling and the EB which arises as a result of coupling between these two layers.

\begin{figure}[t]
	\centering
	\includegraphics[width=0.4\textwidth]{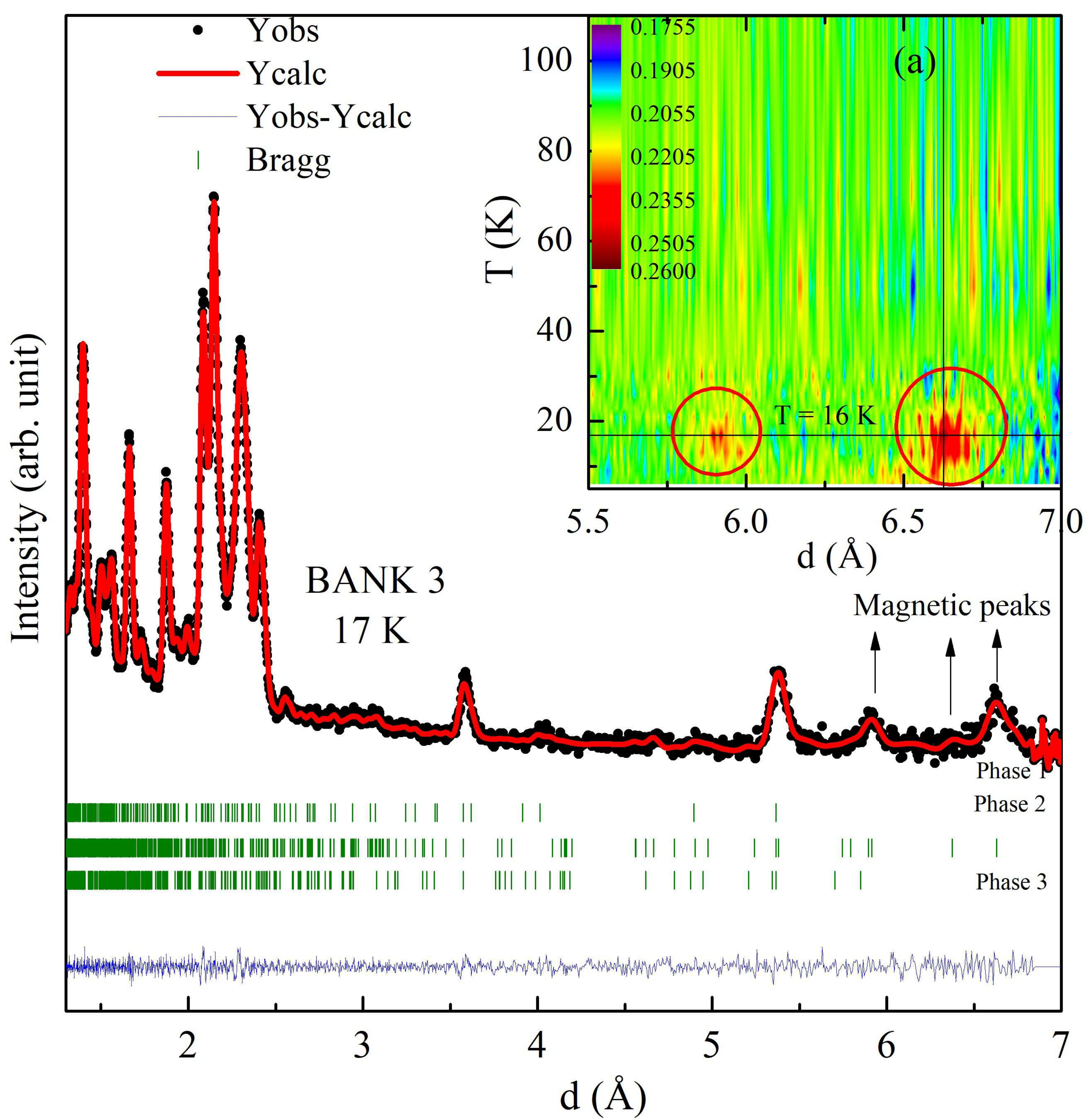}
	\caption{Neutron diffraction pattern of BANK 3 of the GEM diffractometer, measured at 17 K, profile fitted using three phases (see text). Magnetic peaks are indicated by the vertical arrows. Inset show the temperature dependence of magnetic peak intensities, which become diffusive below $\sim$ 16 K.}
	\label{6}
\end{figure}

\begin{figure}[t]
	\centering
	\includegraphics[width=0.42\textwidth]{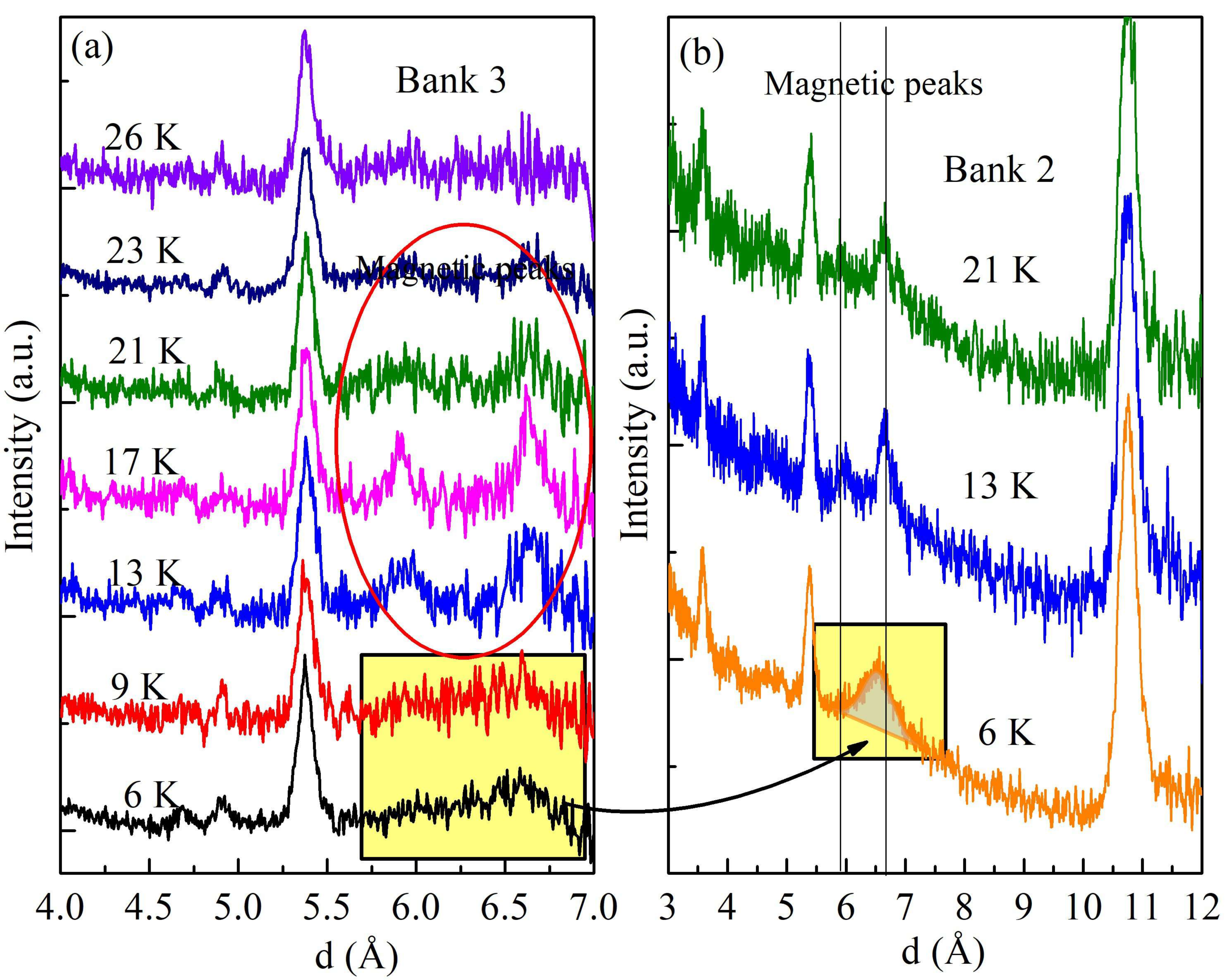}
	\caption{Temperature dependence of selected reflections from neutron diffraction patterns of BANK 3 (a) and BANK 2 (b). At low temperature broadness of magnetic peaks shows the short or glassy magnetic correlations, shown in yellow block.}
	\label{newnpd}
\end{figure}

Fig.~\ref{6} shows the profile matched neutron diffraction pattern measured at 17 K. The crystal structure of CCO can be indexed using superspace group $C2/m(0q0)00$ which is marked as phase 1 having unit cell parameters~\cite{Gre2004} as $a$ = 4.8309 \AA, $b$  = 4.5615 \AA, $c$ = 10.8360 \AA, $\beta$ = 98.134$^\circ$ and $q$ = (0, 1.612, 0). Phases 2 and 3 represent the individual subsystems ([Ca$_2$CoO$_3$] \& [CoO$_2$], respectively), each of them modulated with the magnetic propagation vector, $q_{mag}$ = (0.481, 0.377, 0.0015), obtained using the k-search software~\cite{Rod2001}. Similar three components modulation was observed for the well known SDW material~\cite{And1981} (TMTSF)$_2$ PF$_6$. Phase 2 has lattice parameters as $a$ = 4.8309 \AA, $b_1$  = 4.5615 \AA, $c$ = 10.8360 \AA~and $\beta$ = 98.134$^\circ$ with $q_1$ = $q_{mag}$ = (0.481, 0.377, 0.015), and phase 3 has $a$ = 4.8309 \AA, $b_2$  = 4.5615/$q$ = 2.8297 \AA, $c$ = 10.8360 \AA~and $\beta$ = 98.134$^\circ$ with $q_2$ = (0.481, 0.377*$q$, 0.015). The magnetic modulation is quite complex, as from $C2/m$ symmetry only the $P1$ space group is allowed (found using $MAXMAGN$ program~\cite{Per2015}). From the propagation vector it is clear that the moments are propagating in the $ab$ plane. Interestingly, from the contour plot it is observed that the intensity of magnetic peaks are considerable at $\sim$ 16 K and become diffusive at lower temperatures (see inset (a)). Figs.~\ref{newnpd} (a \& b) show the evolution of magnetic peaks with temperature. As evident from the broadness of magnetic peaks, short range correlations appear below $\sim$ 13 K, which is a direct signature of glassiness~\cite{Ber2016} as observed also in DC magnetization.

\begin{figure}[t]
	\centering
	\includegraphics[width=0.48\textwidth]{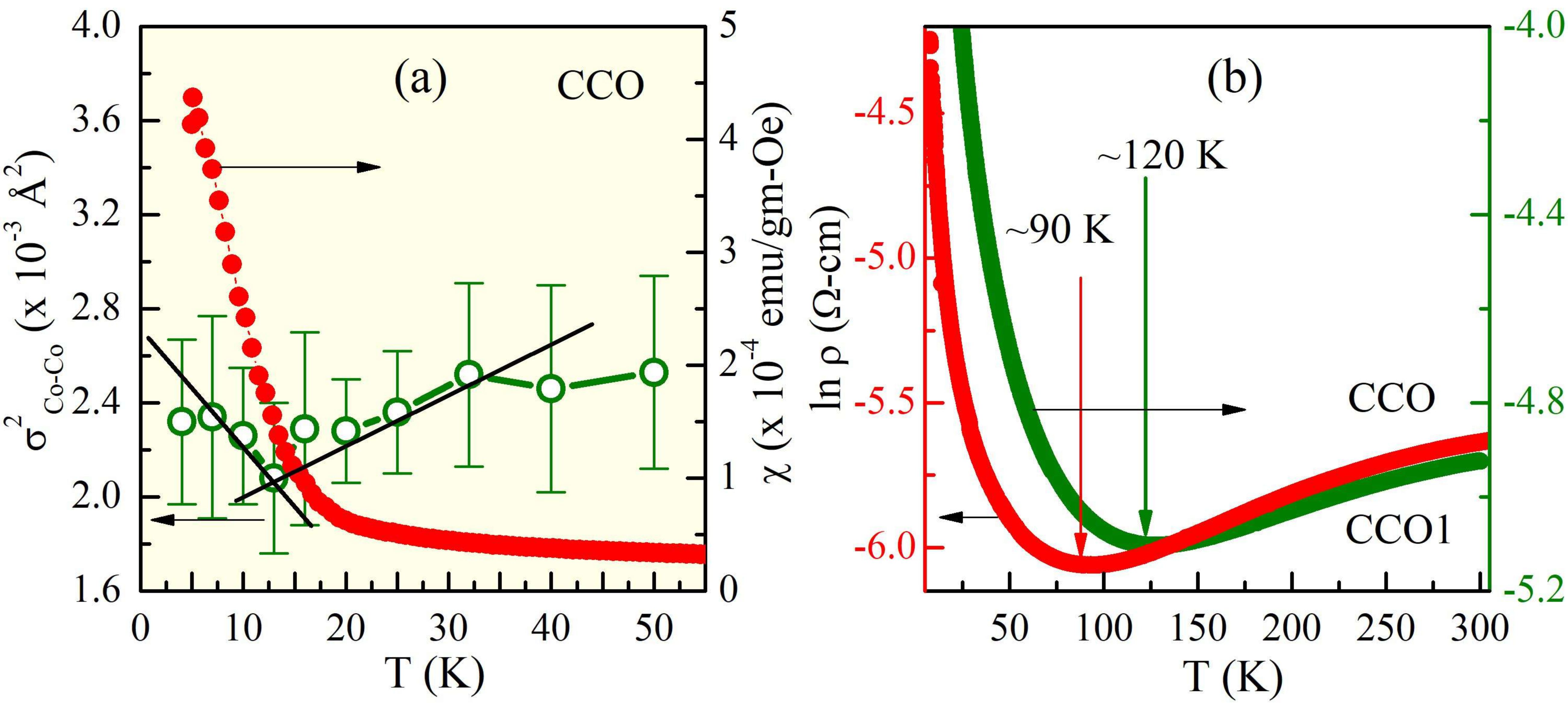}
	\caption{(a) MSRD of Co-Co pair in CoO$_2$ layer $i.e.$, $\sigma^2_{Co-Co}$, showing an anomaly at the susceptibility upturn temperature. Black solid lines are guide to the eye. (b) Resistivity as a function of temperature for CCO and CCO1, showing shift a in T$_{MIT}$ with Tb doping and magnitude change hints towards the change in SDW gap.}
	\label{10}
\end{figure}

It has been reported~\cite{Tys2009} that the Co-Co correlation in the CoO$_2$ results in an anomaly in the mean square relative displacement (MSRD) related to this pair, $i.e.$, $\sigma^2_{Co-Co}$. We have fitted the EXAFS data using a standard protocol~\cite{Teo2012} by assuming the first shell parameters as temperature independent, as observed previously~\cite{Tys2009}, and only iterate the second shell parameters. We observed an anomaly in the $\sigma^2_{Co-Co}$ at temperature $\sim$ 15 K, as shown in Fig.~\ref{10} (a), which matches with the magnetization upturn. This result supports the spin-phonon coupling in the CoO$_2$ layer. This type of observation has been made earlier also by temperature dependent Raman scattering~\cite{An2007}. This type of strong spin-phonon coupling observation and the AFM ordering via neutron diffraction confirm that the AFM SDW is originating from the CoO$_2$ only. Our NPD data clearly show that below 15 K, the SDW in CoO$_2$ tends to become short range and directed along arbitrary directions. This truncated SDW is the reason of glassiness and result into the features of bulk spin glass. The interlayer coupling ($J_\perp$) between CoO$_2$ and Ca$_2$CoO$_3$ becomes stronger below $\sim$ 15 K and is responsible for truncation of SDW in CoO$_2$. The layer of Ca which can control the interlayer coupling and the SDW is altered via Tb doping (having high $c$ axis Ising spins) and hence absence of glassiness in CCO1 ($i.e.$, Tb screens the  effect of rocksalt layer).

To further investigate the effects of Tb substitution, it will be informative to look at the transport results. Fig.~\ref{10} (b) comprises temperature dependent resistivity of both the samples, showing the shift of T$_{MIT}$ towards higher temperature with the Tb doping. The magnitude of resistivity of CCO1 is found larger compared to CCO. We have fitted the curves with the activated behavior (not shown here) using the relation $\rho = \rho_o exp \left(\Delta/k_B T\right)$, as suggested in earlier reports \cite{Sug2002,Mur2017} and found a significant enhancement in the band gap $\Delta$. This shows that the doping in the rocksalt significantly affects the overall band structure and hence also to the SDW gap. The $T_{MIT}$ in general is directly related with the correlation via the relation $k_B T_{MIT} = 1.14~\epsilon_o~e^{\frac{-1}{\lambda_e}}$, where $\lambda_e = U n(E_F)$ is the electron-electron coupling constant~\cite{Gru2018}. Assuming the same density of states $n(E_F)$, for CCO and CCO1, one can see the relatively large $U$ in case of CCO1. It is to be noted that large $U$ links to more localization and less AFM exchange ($J_\parallel~ \propto~ \frac{t^2}{U}$) between spins. Interestingly, this scenario is in accordance with the low value of $\theta_P$ obtained for CCO1 (Fig. S3~\cite{Sup}). However, the above mentioned argument is based on the assumption that both samples (CCO \& CCO1) have the same nesting vector/magnetic vector ($q$) ($i.e.$, same DOS), while in reality the position of the Fermi level controls the $q$.

In conclusion, we have studied the pure and Tb substituted CCO by means of DC magnetization, neutron diffraction, XPS, EXAFS and resistivity measurements. Exchange bias has been observed in both the samples. Glassiness has been found as the origin of larger hysteresis in CCO than in CCO1. Interlayer coupling between triangular (CoO$_2$) and rocksalt (Ca$_2$CoO$_3$) has been attributed as the reason behind ferrimagnetism. Magnetocrystalline anisotropy in the rocksalt layer acts as pinning for EB. Neutron diffraction and EXAFS results combinedly hints that incommensurate SDW is present in the triangular layer, which tends to short ranged below T$_{Ferri}$ and ultimately turn into glassy phase due to stronger
interlayer coupling with cluster ferromagnetism in the rocksalt
layer. Weaker AFM correlation observed in CCO1 is substantiated by increased correlation effects as manifested from electrical transport data, highlighting the intricate relation between magnetism and electron correlations in these samples.

The authors are thankful to Sharad Karwal and A. Wadikar for help during XPS measurements. AA acknowledges UGC, New Delhi for providing financial support in form of the MANF scheme (Grant No. 2016-17/MANF-2015-17-UTT-53853). DKS acknowledges financial support from DST-New Delhi through Grant No. INT/RUS/RFBR/P-269 and SERB-New Delhi through ECR award grant. DST-DESY project is acknowledged for financial support for performing synchrotron EXAFS studies. The work at the ISIS Neutron and Muon facility was supported by the Science and Technology Facilities Council and DST-RAL project. JAHC thanks the Brazilian agencies CNPq and FAPDF for the financial support. 

\bigskip
\bibliography{cco_mag}

\begin{thebibliography}{38}
\bibitem [{Rev()}]{Rev}%
  \BibitemOpen
  \href@noop {} {\bibinfo  {journal} {For Tb$^{3+}$, the free-ion term $^7F_6$
  (L=3, S=3, J=6) is split by crystal field, and the ground state is generally
  formed by two wavefunctions $\arrowvert J, J_z\rangle$ in form of $\arrowvert
  6, 6\rangle$ + $\arrowvert 6, -6\rangle$ and $\arrowvert 6, 6\rangle$ -
  $\arrowvert 6, -6\rangle$ which provide large magnetic moments $\sim$ 9
  $\mu_B$/Tb of Ising-like character, i. e. along $c$-axis.}\ }\BibitemShut
  {NoStop}%
\end{thebibliography}

\end{document}


\title{Supplementary: Magnetic correlations in subsystems of the misfit [Ca$_2$CoO$_3$]$_{0.62}$[CoO$_2$] cobaltate}
\author{Abdul Ahad}
\address{Department of Physics, Aligarh Muslim University, Aligarh 202002, India}
\author{K. Gautam}
\address{UGC-DAE Consortium for Scientific Research, Indore 452001, India}
\author{K. Dey}
\address{UGC-DAE Consortium for Scientific Research, Indore 452001, India}
\author{S. S. Majid}
\thanks{Present address: Optical Physics Lab, Institute of Physics, Academia Sinica, Taipei, Taiwan.}
\address{Department of Physics, Aligarh Muslim University, Aligarh 202002, India}
\author{F. Rahman}
\address{Department of Physics, Aligarh Muslim University, Aligarh 202002, India}
\author{S. K. Sharma}
\address{Department of Physics, Central University of Punjab, Bathinda Punjab 151001, India}
\author{J. A. H. Coaquira}
\address{LSNCM-NFA, Institute of Physics, UnB, Brasilia DF 70910 900, Brazil}
\author{Ivan da Silva}
\address{ISIS Facility, Rutherford Appleton Laboratory, Chilton, Didcot OX11 0QX, United Kingdom}
\author{E. Welter}
\address{Deutsches Elektronen-Synchrotron, Notkestrasse 85, D-22607 Hamburg, Germany}
\author{D. K. Shukla}
\thanks{Corresponding Author: dkshukla@csr.res.in}
\address{UGC-DAE Consortium for Scientific Research, Indore 452001, India}


\date{\today}

 \maketitle

\newpage

\section{Magnetic entropy}

\begin{figure}[hbt]
	\centering
	\includegraphics[width=0.9\textwidth]{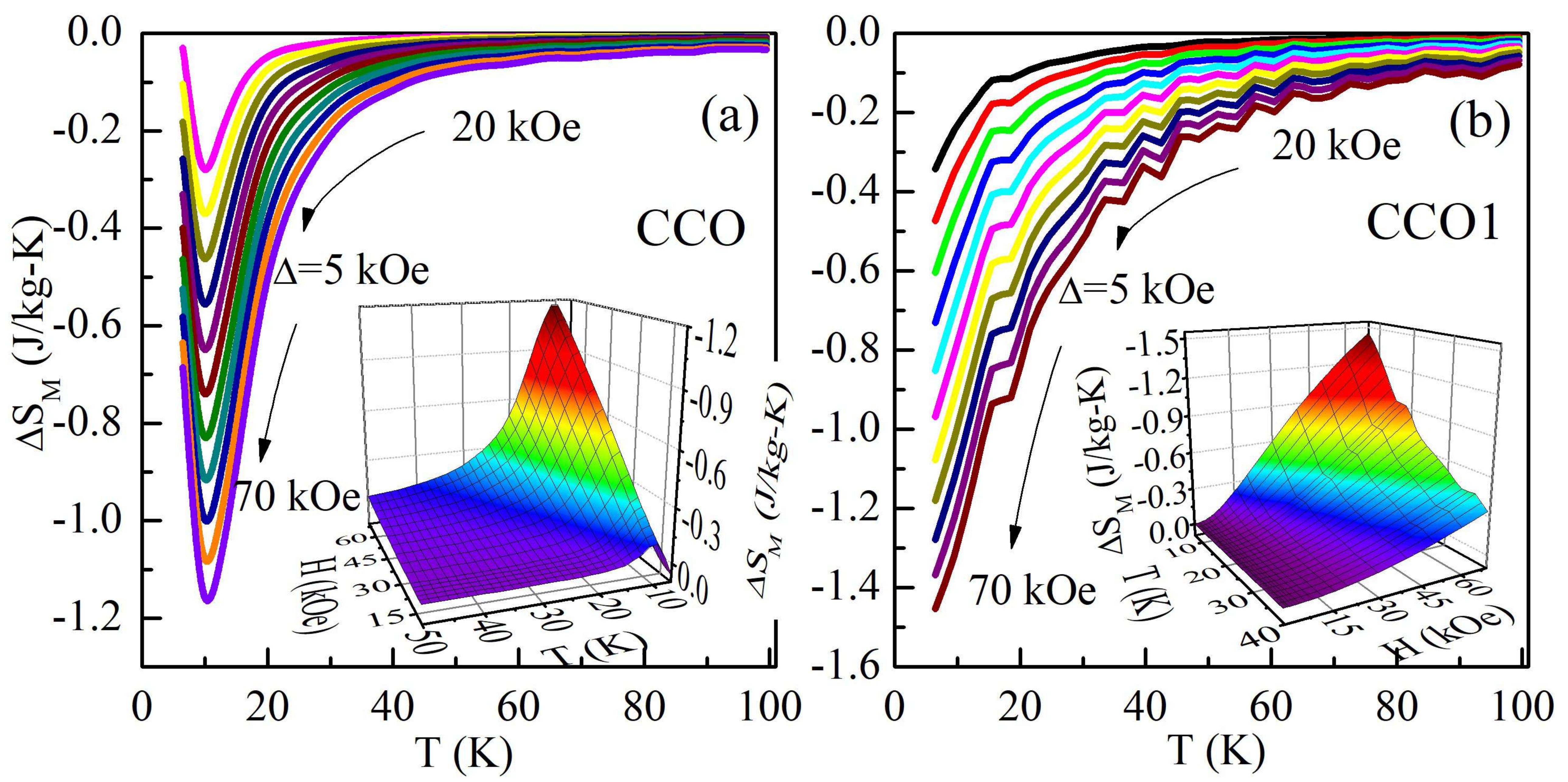}
	\caption{Temperature dependent $\Delta S_M$ for different fields. Inset shows the 3D contour plot ($\Delta S_M$-H-T) (a) for CCO and (b) for CCO1.}
	\label{3a}
\end{figure}

Fig.~\ref{3a} (a \& b) displays the change in magnetic entropy ($\Delta S_M$) with respect to temperature, for CCO and CCO1, respectively, calculated for different applied fields using Maxwell's equation, $\Delta S_M = \int^b_a \left(\frac{dM}{dT}\right)_H dH$. A transition is found at $\sim$ 10 K which we denote as the T$_{Ferri}$. This reflects the first derivative of M (T) while in literature T$_{Ferri}$ was calculated by the upturn of M(T). For CCO1, no such transition observed in $\Delta S_M$.

\cleardoublepage

\section{Arrott plots}
\begin{figure}[hbt]
	\centering
	\includegraphics[width=0.9\textwidth]{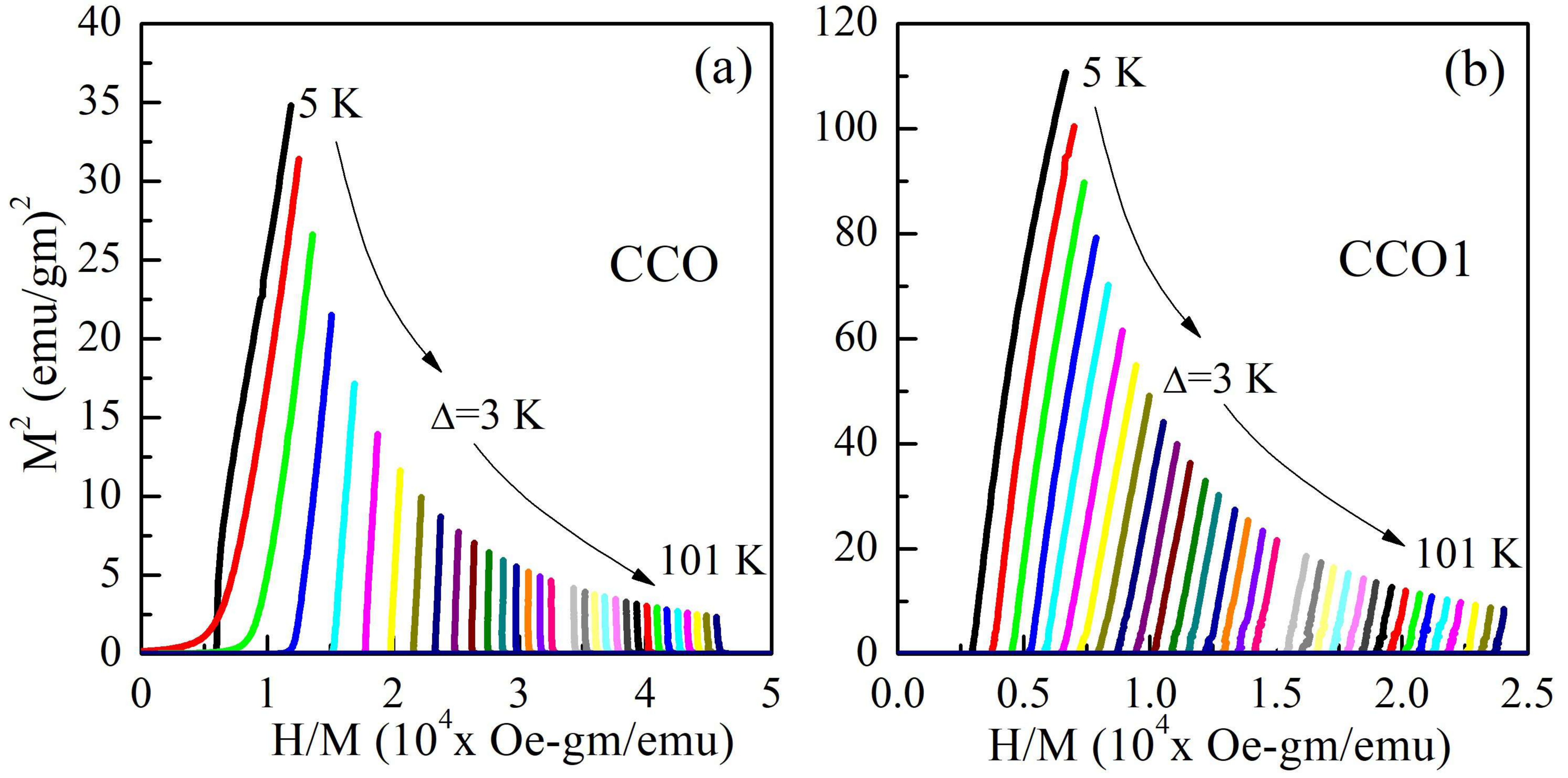}
	\caption{Arrott plot between 5 to 101 K showing absence of spontaneous magnetization (a) for CCO, and (b) for CCO1.}
	\label{2a}
\end{figure}

Fig.~\ref{2a} (a \& b) shows the Arrott plot (5 to 101 K), representing absence of spontaneous magnetization (no intercept at y-axis) in both, CCO and CCO1.

\cleardoublepage

\section{Curie-Weiss fitting}
Curie-Weiss fitting to $\chi^{-1}$ vs T data measured at 10 Oe shows the negative Weiss temperature ($\theta_p$) indicating the AFM correlations but weaker in case of CCO1 than in CCO.

\begin{figure}[hbt]
	\centering
	\includegraphics[width=0.9\textwidth]{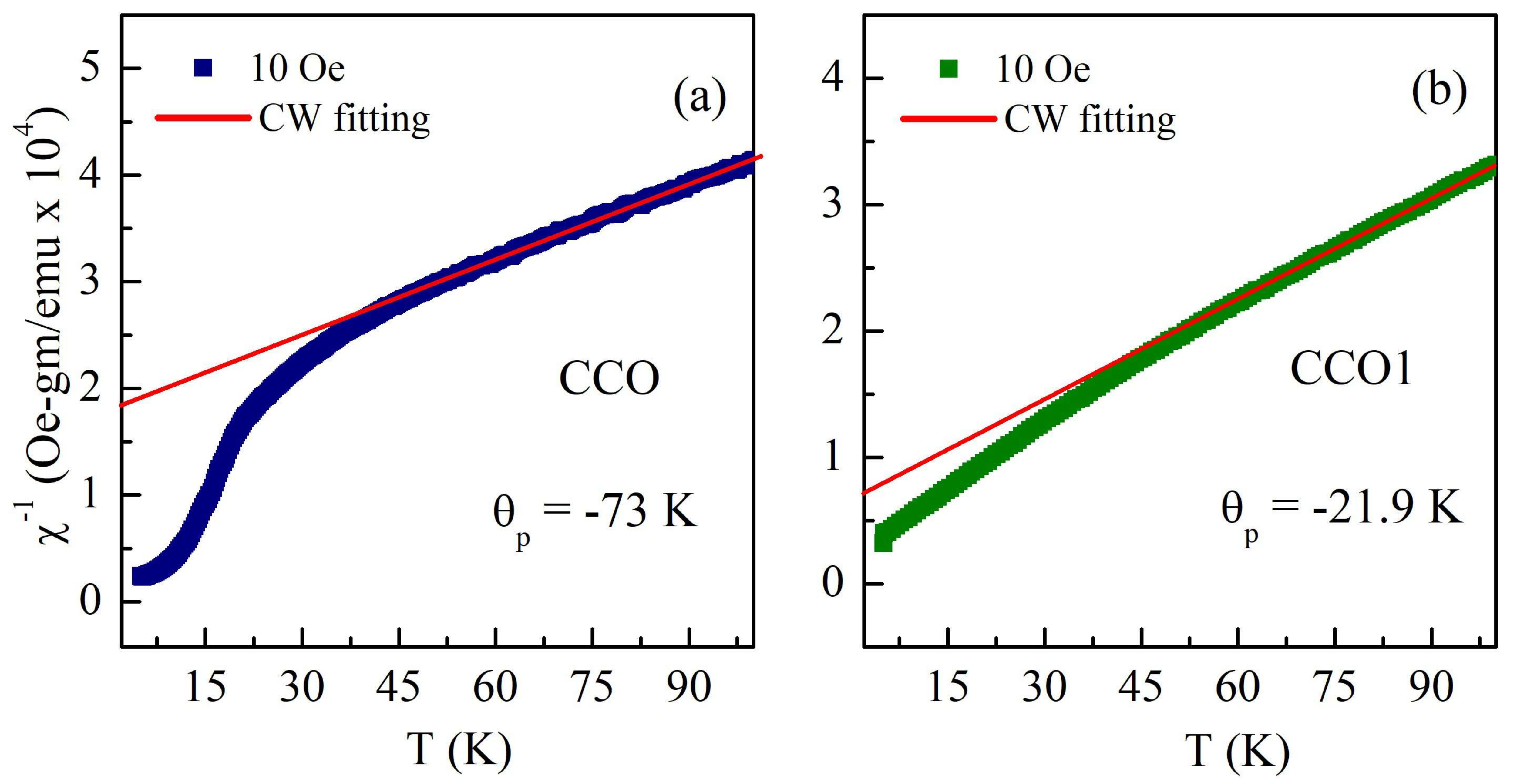}
	\caption{Curie-Weiss fits to the temperature dependent inverse susceptibility ($\chi^{-1}$(T)) data of (a) CCO and (b) CCO1.}
	\label{7b}
\end{figure}

\cleardoublepage

\section{X-ray photoemission spectroscopy}
\begin{figure}[hbt]
	\centering
	\includegraphics[width=0.8\textwidth]{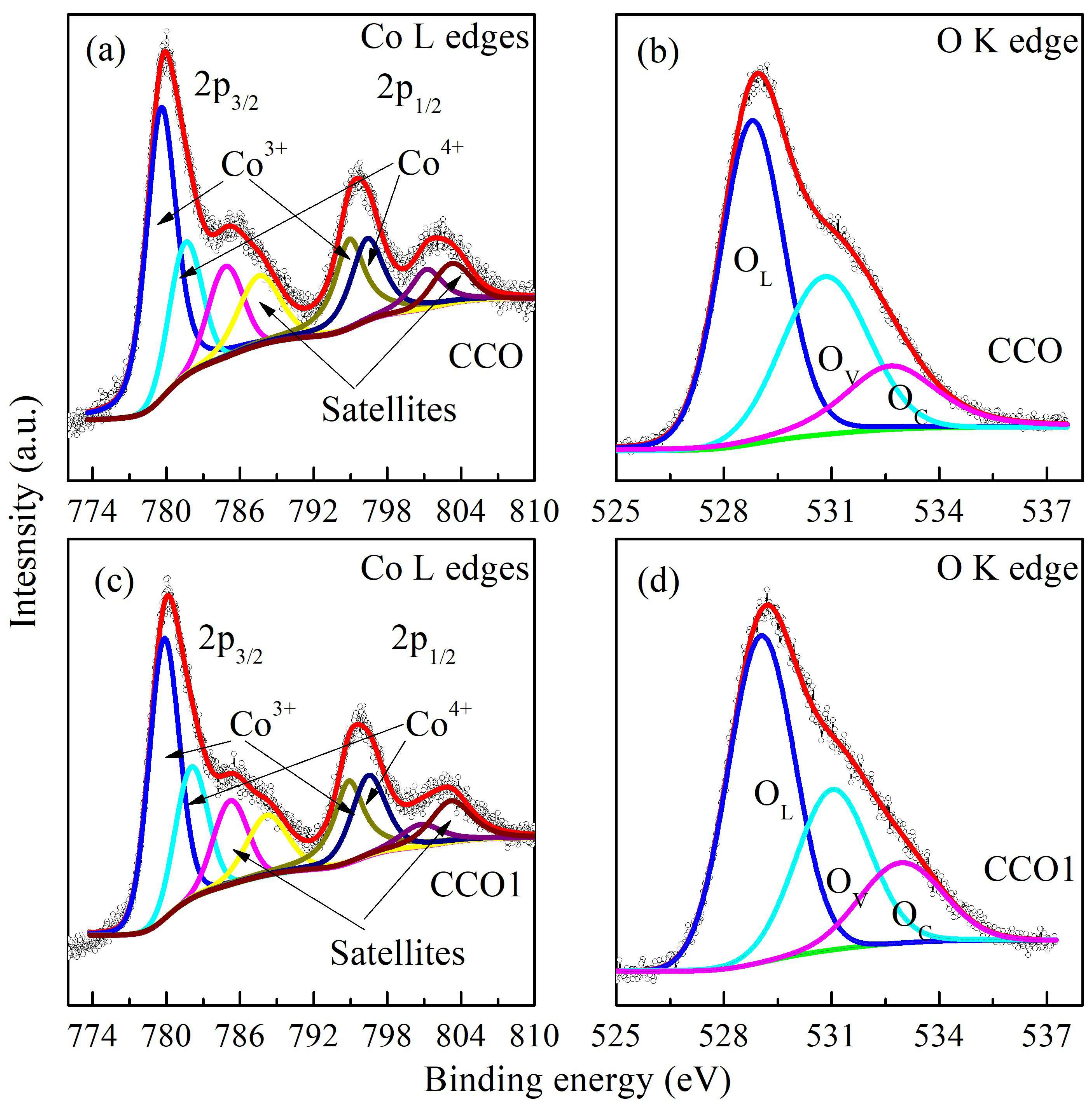}
	\caption{(a \& b) Co 2p and O 1s XPS spectra showing the mixed valency of Co (+3 \& +4) and oxygen vacancy in the lattice, for CCO (c \& d) for CCO1.}
	\label{9}
\end{figure}

 Fig.~\ref{9} (a \& b) display the Co 2p and O 1s XPS results for CCO and (c \& d) show the corresponding for CCO1. The fraction of Co$^{3+}$ (calculated from the peak ratios) in CCO is $\sim$ 68 \% while for CCO1 it is $\sim$ 69.5 \%.

